\documentclass[article]{IEEEtran}
\usepackage{color}
\usepackage{threeparttable}
\usepackage{srcltx}

\usepackage{cite,graphicx,amsmath,psfrag,bm}
\usepackage[usenames,dvipsnames]{xcolor}
\usepackage{mathabx}
\usepackage{subfig}
\usepackage{cancel}
\usepackage{epstopdf}
\usepackage{pstool}

\begin{document}

\title{Dielectric Resonator Metasurface \\for Dispersion Engineering}
\author{Karim~Achouri,~\IEEEmembership{Student Member,~IEEE,} Ali~Yahyaoui, Shulabh~Gupta,~\IEEEmembership{Member,~IEEE,} \\
        Hatem~Rmili,~\IEEEmembership{Member,~IEEE,} and~Christophe~Caloz,~\IEEEmembership{Fellow,~IEEE}% <-this % stops a space
\thanks{K. Achouri and C. Caloz are with the Department of Electrical Engineering, Polytechnique Montreal, Montreal, Quebec, Canada email: karim.achouri@polymtl.ca, christophe.caloz@polymtl.ca}% <-this % stops a space
\thanks{S. Gupta is with the Department
of Electronics, Carleton University, Ottawa, Ontario, Canada e-mail: shulabh.gupta@carleton.ca}% <-this % stops a space
\thanks{A. Yahyaoui and H. Rmili are with the Department
of Electrical and Computer Engineering, King Abdulaziz University, Jeddah, Saudi Arabia e-mail: hmrmili@kau.edu.sa. They are also with
Sys-Com, ENIT, University of Tunis El Manar, BP 37, Belvedere 1002 Tunis, Tunisia.}% <-this % stops a space
}

%\markboth{IEEE TRANSACTIONS ON ANTENNAS AND PROPAGATION, VOL. X, NO. Y, MONTH Z}%

\maketitle

\begin{abstract}
We introduce a practical dielectric metasurface design for microwave frequencies. The metasurface is made of an array of dielectric resonators held together by dielectric connections thus avoiding the need of a mechanical support in the form of a dielectric slab and the spurious multiple reflections that such a slab would generate. The proposed design can be used either for broadband metasurface applications or monochromatic wave transformations. The capabilities of the concept to manipulate the transmission phase and amplitude of the metasurface are supported by numerical and experimental results. Finally, a half-wave plate and a quarter-wave plate have been realized with the proposed concept.
\end{abstract}

\begin{IEEEkeywords}
Dispersion engineering, electromagnetic metamaterials, metasurfaces, ultrafast electronics, ultrafast optics.
\end{IEEEkeywords}

%\tableofcontents

\section{Introduction}

Metasurfaces, the two-dimensional counterparts of metamaterials, allow the manipulation of electromagnetic waves via adequate engineering of their scattering particles. Due to their unprecedented capabilities of broadband spatial wavefront manipulations~\cite{doi:10.1021/nl402039y}, they have already led to multiple applications involving dispersion compensation~\cite{Achromatic_MS_Capasso,Kivshar_Alldielectric}, reflection-less sheets~\cite{BBParticlesTratyakov} and quarter and half-wave plates~\cite{BB_PLates_MS}.

The concept of real-time analog signal processing (R-ASP), based on dispersion engineering, consists in manipulating signals in time rather than in space~\cite{Jour:2013_MwMag_Caloz,Dragoman_THZ}. However, most R-ASP systems reported to date have been restricted to guided-wave components, as is for instance the case with phasers~\cite{Phaser_Def_Gupta}. This suggests that broadband metasurfaces, beside their spatial wavefront control capabilities, could play an important role as dispersion based real-time signal processors, where they could act as spatial phasers, as suggested in~\cite{Gupta_AllDMS_CAMA,AchouriEPJAM}.

While most microwave metasurfaces have been realized with metallic scattering particles~\cite{AchouriEPJAM}, all-dielectric implementations of metasurfaces are particularly attractive due to their low loss and ease of fabrication. Relatively few all-dielectric metasurfaces have been reported so far and they generally operate in the optical regime~\cite{Kivshar_Alldielectric,DMS_EIT,Tuned_DMS,Elliptical_DMS,Cylindrical_DMS,Shalav_Alldielectric}.

In this work, we propose an experimental demonstration of a broadband all-dielectric metasurface design at microwave frequencies \cite{Gupta_AllDMS_CAMA}. This new metasurface structure is very simple to fabricate using high-precision laser cutting. The proposed structure may be used for real-time analog signal processing, as discussed in~\cite{Gupta_AllDMS_CAMA,AchouriEPJAM}, as well as for monochromatic wave transformations. As an example of the proposed design, we present the implementation of two metasurfaces respectively performing the operations of half-wave and quarter-wave plates.

This paper is organized as follows: the general concept of dielectric metasurfaces is discussed and the proposed microwave implementation is introduced in the next section. Then, in the third section, the dispersion response of such metasurfaces is mathematically analysed and demonstrated by numerical simulations and experimental measurements. Finally, in the last section, two monochromatic dielectric metasurfaces performing the operation of half- and quarter-wave plates are presented.

\section{Dielectric Resonator Structure}

\subsection{Ideal Design}

\begin{figure}[htbp]
\begin{center}
\includegraphics[width=\columnwidth]{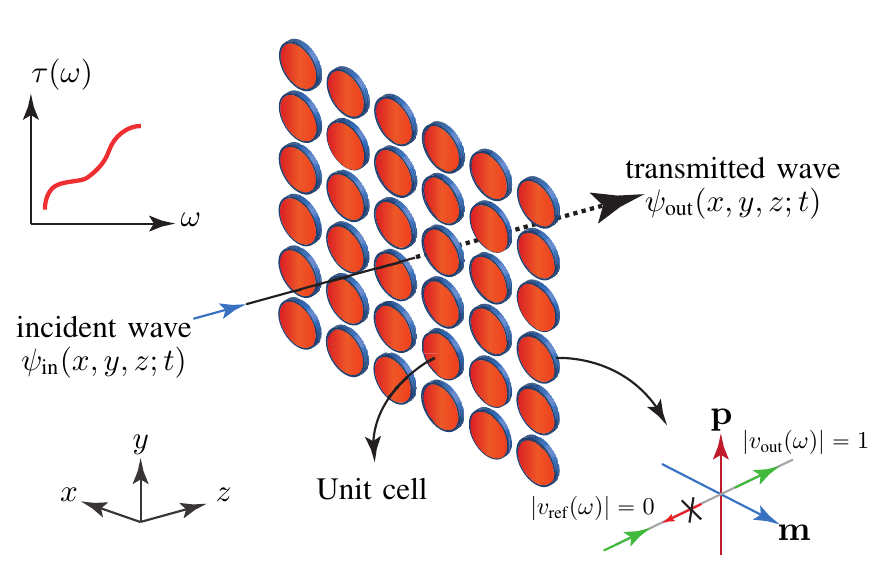}
%\psfragfig*[width=\columnwidth]{Phaser_Metasurface}{
%\psfrag{h}[c][c][0.9]{\shortstack{incident wave\\$\psi_\text{in}(x,y, z; t)$}}
%\psfrag{c}[c][c][0.9]{\shortstack{transmitted wave\\$\psi_\text{out}(x,y, z; t)$}}
%\psfrag{x}[c][c][0.9]{$x$}
%\psfrag{y}[c][c][0.9]{$y$}
%\psfrag{z}[c][c][0.9]{$z$}
%\psfrag{e}[l][l][0.9]{$\tau(\omega)$}
%\psfrag{f}[c][c][0.9]{$\omega$}
%\psfrag{d}[c][c][0.9]{Unit cell}
%\psfrag{A}[c][c][1]{$\mathbf{p}$}
%\psfrag{B}[c][c][1]{$\mathbf{m}$}
%\psfrag{C}[c][c][0.7]{$|v_\text{out}(\omega)|=1$}
%\psfrag{D}[c][c][0.7]{$|v_\text{ref}(\omega)|=0$}}
\caption{All-dielectric metasurface, formed by a 2D array of dielectric resonators.}
\label{Fig:SpatialPhaser}
\end{center}
\end{figure}

A dielectric resonator metasurface consists in a two-dimensional periodic array of dielectric resonators, as depicted in Fig.~\ref{Fig:SpatialPhaser}. The thickness of the metasurface and the period of the unit cells in the array are both sub-wavelength. The metasurface can be uniform (made of a repetition of identical resonators), or non-uniform if electromagnetic transformations, such as for instance generalized refraction, are required. The dielectric resonators may exhibit 90$^\circ$ symmetry in the plane of the metasurface ($x-y$ plane), or be asymmetric to allow independent control of the $x$ and $y$ polarizations. As will be shown shortly, the electromagnetic response of such a metasurface is, due to the Lorentzian behavior of the dielectric resonators, inherently broadband, and may be controlled by tuning the response of each resonator. The capability to control the group delay, as depicted in Fig.~\ref{Fig:SpatialPhaser}, in order to realize real-time processors may be achieved by cascading several dielectric metasurfaces with different with resonators exhibiting different Lorentzian responses~\cite{Gupta_AllDMS_CAMA,AchouriEPJAM}.
%
%\begin{figure}[htbp]
%\begin{center}
%\includegraphics[width=\columnwidth]{Figures/Phaser_Metasurface}
%%\psfragfig*[width=\columnwidth]{Figures/Phaser_Metasurface}{
%%\psfrag{h}[c][c][0.9]{\shortstack{incident wave\\$\psi_\text{in}(x,y, z; t)$}}
%%\psfrag{c}[c][c][0.9]{\shortstack{transmitted wave\\$\psi_\text{out}(x,y, z; t)$}}
%%\psfrag{t}[c][c][0.9]{$t$}
%%\psfrag{a}[c][c][0.9]{$\omega_0$}
%%\psfrag{b}[c][c][0.9]{$\omega(t)$}
%%\psfrag{x}[c][c][0.9]{$x$}
%%\psfrag{y}[c][c][0.9]{$y$}
%%\psfrag{z}[c][c][0.9]{$z$}
%%\psfrag{e}[c][c][0.9]{$\tau(\omega)$}
%%\psfrag{f}[c][c][0.9]{$\omega$}
%%\psfrag{d}[c][c][0.9]{Unit cell}
%%\psfrag{A}[c][c][1]{$\mathbf{p}$}
%%\psfrag{B}[c][c][1]{$\mathbf{m}$}
%%\psfrag{C}[c][c][0.7]{$|v_\text{out}(\omega)|=1$}
%%\psfrag{D}[c][c][0.7]{$|v_\text{ref}(\omega)|=0$}}
%\caption{A spatial phaser formed using a 2D array of dielectric resonators, chirping an incident pulsed wavefront according to the specified dispersion resonse.}
%\label{Fig:SpatialPhaser}
%\end{center}
%\end{figure}
%

\subsection{Microwave Design}

\begin{figure}[h!]
\begin{center}
\includegraphics[width=\columnwidth]{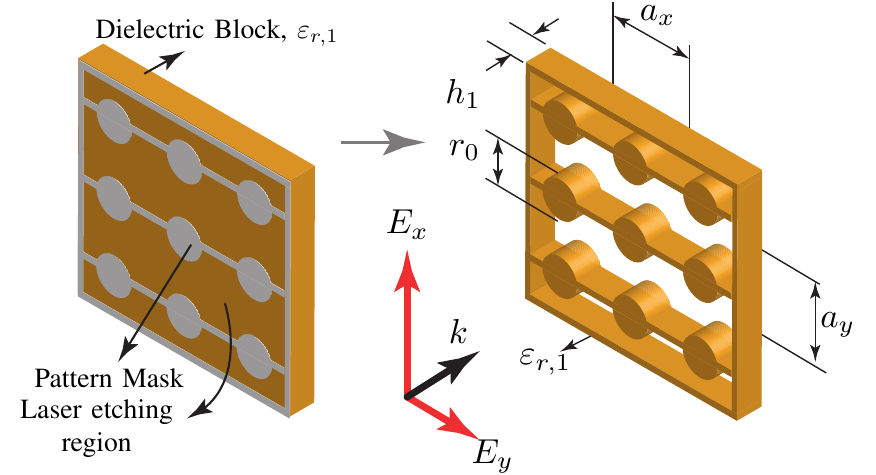}
%\psfragfig*[width=\columnwidth]{IC_APMS}{
%\psfrag{a}[c][c][1]{$a_x$}
%\psfrag{b}[c][c][1]{$a_y$}
%\psfrag{c}[c][c][1]{$\varepsilon_{r,1}$}
%\psfrag{e}[c][c][1]{$r_0$}
%\psfrag{f}[c][c][1]{$h_1$}
%\psfrag{g}[c][c][1]{$h_2$}
%\psfrag{h}[c][c][1]{$E_x$}
%\psfrag{i}[c][c][1]{$k$}
%\psfrag{j}[c][c][1]{$E_y$}
%\psfrag{m}[c][c][0.8]{Dielectric Block, $\varepsilon_{r,1}$}
%\psfrag{k}[c][c][0.8]{Pattern Mask}
%\psfrag{n}[r][c][0.8]{\shortstack{Laser etching\\ region}}}
\caption{Proposed configuration for an all-dielectric metasurface formed by an array of interconnected cylindrical dielectric resonators.}
\label{Fig:FabTech}
\end{center}
\end{figure}

In the optical regime, dielectric metasurfaces typically consist in dielectric resonators placed on dielectric slab (usually silica) for mechanical support~\cite{DMS_EIT,Tuned_DMS,Elliptical_DMS,Cylindrical_DMS,Shalav_Alldielectric}. In this work, we propose an alternative design for microwave dielectric metasurfaces. The design, presented in Fig.~\ref{Fig:FabTech}, consists in resonators mechanically held together by thin dielectric connections. The structure is made by laser cutting a dielectric slab with the appropriate pattern mask, as illustrated in Fig.~\ref{Fig:FabTech}. In addition to being easy to realize, this design has the advantage of avoiding the issue of multiple reflections inside the supporting dielectric slab, which is inevitable in optical dielectric metasurfaces. To achieve the required thickness, several dielectric slabs may be glued together, without significant alteration in the response of the final structure. The supporting connections have a very small width compared to the wavelength and their presence is negligible for the polarization perpendicular to them ($E_x$ in Fig.~\ref{Fig:FabTech}). For the other polarization, the disturbance effect of these connections is small (but not negligible) and can be compensated for by optimization.

\begin{figure}[h!]
\begin{center}
\includegraphics[width=0.9\columnwidth]{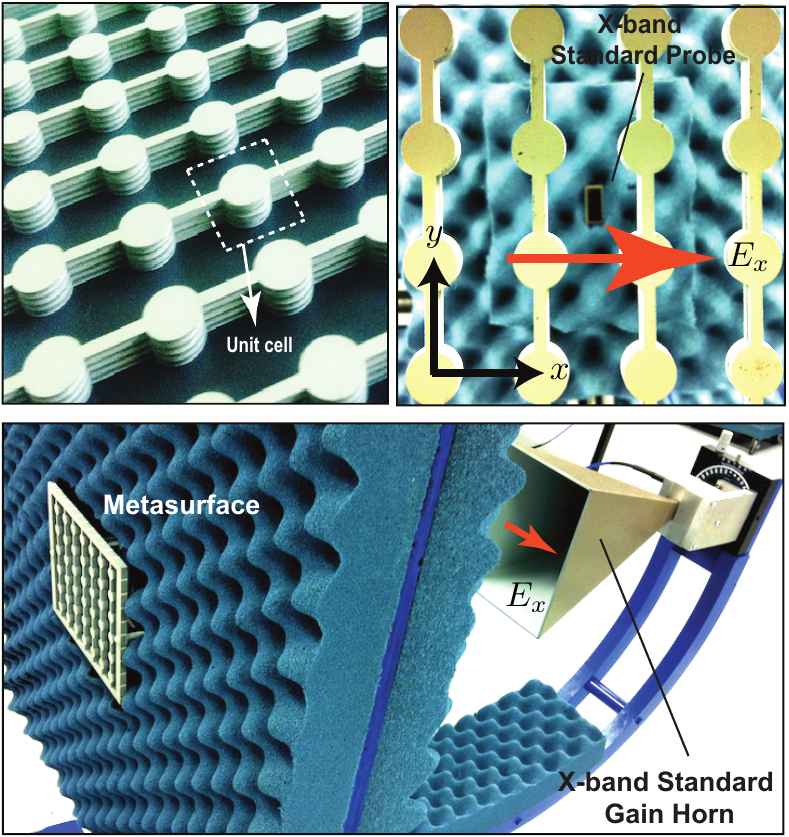}
%\psfragfig*[width=0.9\columnwidth]{Prototype}{
%\psfrag{a}[c][c][1]{$E_{x}$}
%\psfrag{x}[c][c][1]{$x$}
%\psfrag{y}[c][c][1]{$y$}}
\caption{Metasurface made with interconnected dielectric resonators and measurement setup.}
\label{Fig:Setup}
\end{center}
\end{figure}

Figure~\ref{Fig:Setup} shows a fabricated metasurface made by sandwiching several patterned dielectric slabs glued together. It also shows the metasurface measurement setup used to characterize the structures. A gain horn antenna is used to illuminate the metasurface while a standard probe scans a transverse plane in the transmission side of the metasurface.

\section{Matched Dispersion Response}

\subsection{Mathematical Analysis}

The electromagnetic response of a dielectric resonator metasurface can be evaluated by considering the metasurface as an array of orthogonal electric ($\mathbf{p}$) and magnetic ($\mathbf{m}$) dipoles~\cite{Kerker_Scattering}, as suggested in Fig.~\ref{Fig:SpatialPhaser}. It is assumed that the metasurface is perfectly uniform and that the resonators are not held together by dielectric connections. If the metasurface is illuminated by a normally incident plane wave, it generates a normally reflected plane wave and a normally transmitted plane wave with reflection and transmission coefficients given by~\cite{Metasurface_Synthesis_Caloz}
\begin{subequations}
\begin{equation}
R(\omega)=\frac{2jk_0\left(\chi_{\text{mm}}-\chi_{\text{ee}}\right)}{\left(2+jk_0\chi_{\text{ee}}\right)\left(2+jk_0\chi_{\text{mm}}\right)},
\label{Eq:Romega}
\end{equation}
\begin{equation}
T(\omega)=\frac{4+\chi_{\text{ee}}\chi_{\text{mm}}k_0^2}{(2+jk_0\chi_{\text{ee}})(2+jk_0\chi_{\text{mm}})},
\label{Eq:Tomega}
\end{equation}
\label{Eq:RT}
\end{subequations}
where $k_0$ is the wavenumber in free space, and $\chi_{\text{ee}}$ and $\chi_{\text{mm}}$ are the electric and magnetic susceptibilities of the metasurface, respectively, and are a function of $\omega$. The condition for perfect matching ($R(\omega) =0$) is read out from~\eqref{Eq:Romega} as $\chi_\text{mm} = \chi_\text{ee}$. In this case, the transmission function~\eqref{Eq:Tomega} reduces to
\begin{equation}
T(\omega) = \frac{1 - jk_0\chi_{\text{ee}}/2}{1 + jk_0\chi_{\text{ee}}/2},\label{Eq:Trans}
\end{equation}
which, for a special case of zero loss, takes the standard form of an all-pass response~\cite{Cristal_TMTT_01_1969}. Let us now assume a Lorentzian dispersion profile for modelling the susceptibility of a physical system. The corresponding susceptibility function reads
\begin{equation}
\chi_{\text{ee}}(\omega)  = \frac{2A\omega_p^2}{(\omega_0^2 - \omega^2) +j\gamma\omega},\label{Eq:LorentzFit}
\end{equation}
where $A$ is the amplitude, $\gamma$ is a damping factor, $\omega_0$ is the resonance frequency and $\omega_p$ is the plasma frequency. Substituting~\eqref{Eq:LorentzFit} into~\eqref{Eq:Trans} transforms the metasurface transfer function to
\begin{equation}
T(\omega)  = \frac{(\omega_0^2 - \omega^2)-j (Ak_0\omega_p^{2} - \omega\gamma)}{(\omega_0^2 - \omega^2)+j (Ak_0\omega_p^{2} + \omega\gamma)}.\label{eq:Trans_disp}
\end{equation}

\begin{figure}[htbp]
\begin{center}
\includegraphics[width=\columnwidth]{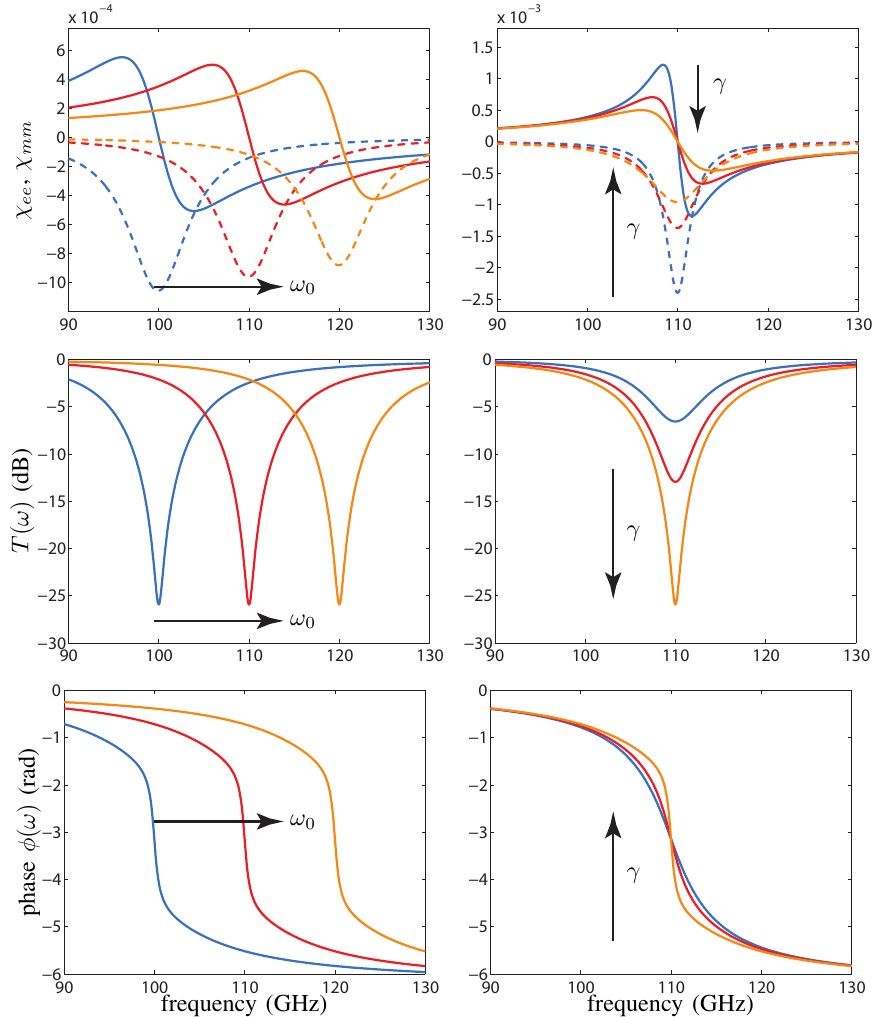}
%\psfragfig*[width=\columnwidth]{AllFig}{
%\psfrag{x}[c][c][0.7]{frequency (GHz)}
%\psfrag{c}[c][c][0.7]{$\chi_{ee}$, $\chi_{mm}$}
%\psfrag{d}[c][c][0.7]{$T(\omega)$ (dB)}
%\psfrag{e}[c][c][0.7]{phase $\phi(\omega)$ (rad)}
%\psfrag{a}[c][c][0.7]{$\omega_0$}
%\psfrag{b}[c][c][0.7]{$\gamma$}}
\caption{Effect of the resonant frequency and loss on the magnitude and phase transmission of a metasurface formed by a 2D square array of dispersion-matched Lorentz oscillators. Left column, varying resonant frequencies ($\gamma = 50\times10^9$, $\omega_0 =2\pi [100\; 110\;120]\times 10^9$), and, right column, varying loss coefficient $\gamma$ ($\gamma = [20\;35\;50]\times10^9$, $\omega_0 =2\pi 110\times 10^9$). In all cases, $A = 10^{-5}$ and $\omega_p=2\pi205\times10^9$}
\label{Fig:LO_Parametric}
\end{center}
\end{figure}

A close inspection of this relation reveals that, interestingly, a lossless ($\gamma = 0$) array of dielectric resonators with equalized electric and magnetic dispersions ($\chi_\text{ee}(\omega) = \chi_\text{mm}(\omega)$) exhibits a flat and broadband transmission ($|T(\omega)|=1$), while presenting a $2\pi$ phase shift around the resonance frequency. Typical transmission and dispersion responses for different resonant frequencies and losses are plotted in Fig.~\ref{Fig:LO_Parametric}.
The real (solid lines) and imaginary (dashed lines) parts, respectively, of the equalized electric and magnetic susceptibilities are shown in the two first figures. Then, the frequency dependent transmission is shown and, as can be seen, only one dip is visible and is directly proportional to the absorption coefficient, $\gamma$. Finally, the transmission phase is plotted. The phase undergoes a $2\pi$ variation and while the phase range remains the same, the inflexion point becomes sharper and sharper when $\gamma$ increases.

\subsection{Simulation \& Experimental Demonstrations}

In order to evaluate the validity of the model developed above, we have simulated and fabricated dielectric metasurfaces and have compared their real responses to the previously discussed ideal responses. The first simulations consist in a metasurface that is similar to the ideal one shown in Fig.~\ref{Fig:SpatialPhaser} in the sense that no dielectric connections have been used. The dielectric resonators have the permittivity $\epsilon_r = 6.15$ (Rogers RO3006), thickness $t=1.27$~mm, radius $r=1.6$~mm and unit cell period $P = 4.25$~mm.
\begin{figure}[h!]
\begin{center}
\subfloat[]{
\includegraphics[width=1\columnwidth]{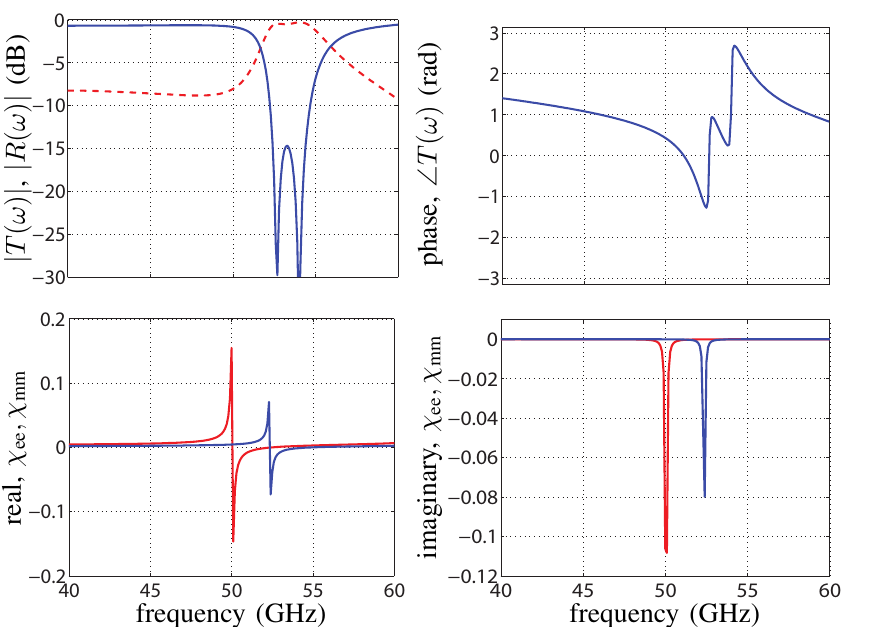}
%\psfragfig*[width=1\columnwidth]{UnBalSX}{
%\psfrag{x}[c][b][0.8]{frequency (GHz)}
%\psfrag{a}[c][c][0.8]{$|T(\omega)|$, $|R(\omega)|$ (dB)}
%\psfrag{b}[c][c][0.8]{phase, $\angle T(\omega)$ (rad)}
%\psfrag{c}[c][c][0.8]{real, $\chi_\text{ee}, \chi_\text{mm}$}
%\psfrag{d}[c][c][0.8]{imaginary, $\chi_\text{ee}, \chi_\text{mm}$}}
\label{Fig:PuckFieldsUnBal1}}\\
\subfloat[]{
\includegraphics[width=0.9\columnwidth]{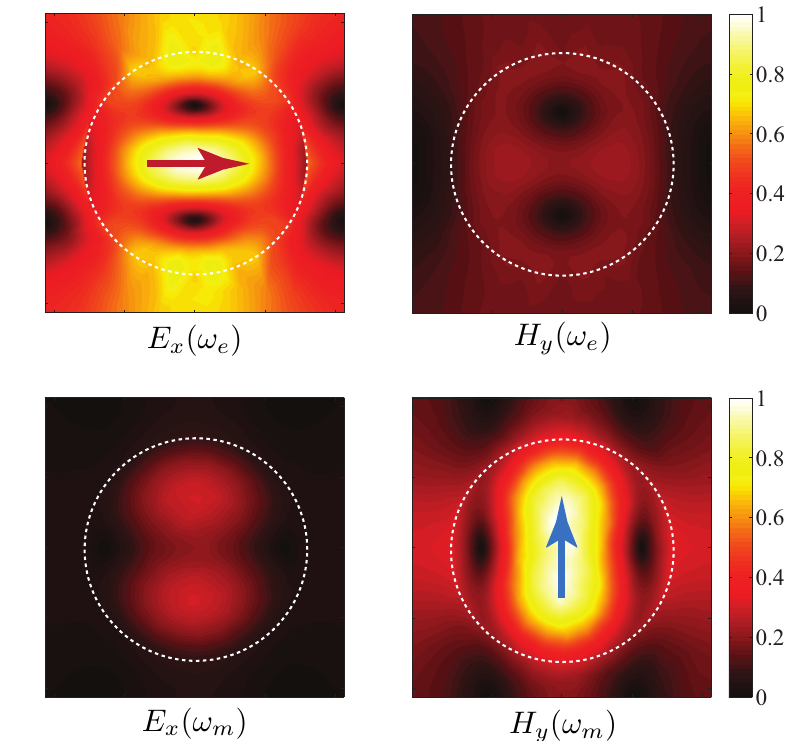}
%\psfragfig*[width=0.9\columnwidth]{UnBalFields}{
%\psfrag{a}[c][c][0.9]{${E_x}(\omega_{e})$}
%\psfrag{c}[c][c][0.9]{${H_y}(\omega_{m})$}
%\psfrag{d}[c][c][0.9]{${E_x}(\omega_{m})$}
%\psfrag{b}[c][c][0.9]{${H_y}(\omega_{e})$}}
\label{Fig:PuckFieldsUnBal2}}
\caption{Simulated unmatched metasurface. (a)~Top, transmission (solid blue line) and reflection (dashed red line) coefficients on the left and, transmission phase on the right. Bottom, real part of the retrieved electric (red line) and magnetic (blue line) susceptibilities on the left and corresponding imaginary parts on the right. (b) Electric and magnetic field distributions in the electric and magnetic resonances, respectively. All results are computed using HFSS.}
\label{Fig:PuckFieldsUnBal}
\end{center}
\end{figure}
\\

\begin{figure}[h!]
\begin{center}
\subfloat[]{
\includegraphics[width=1\columnwidth]{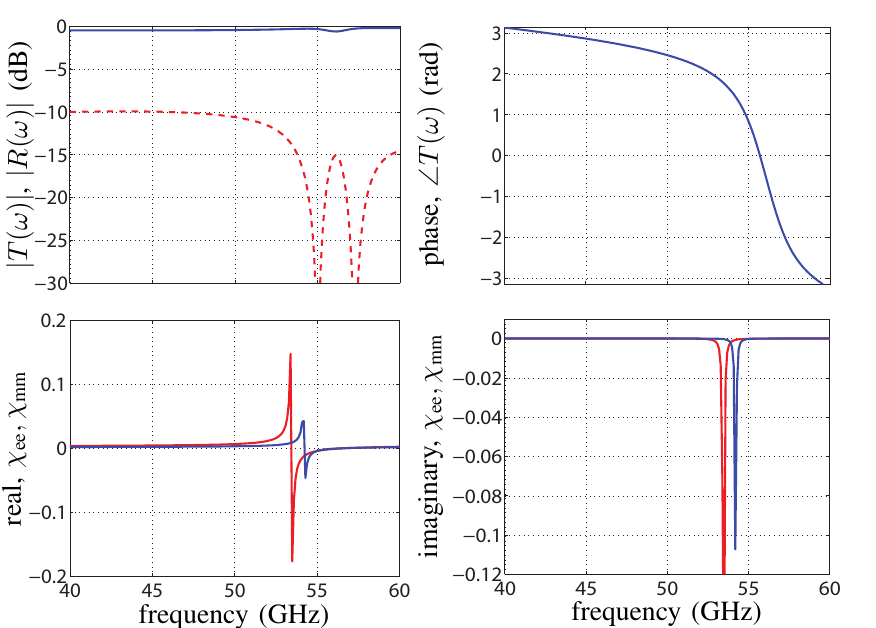}
%\psfragfig*[width=1\columnwidth]{BalSX}{
%\psfrag{x}[c][b][0.8]{frequency (GHz)}
%\psfrag{a}[c][c][0.8]{$|T(\omega)|$, $|R(\omega)|$ (dB)}
%\psfrag{b}[c][c][0.8]{phase, $\angle T(\omega)$ (rad)}
%\psfrag{c}[c][c][0.8]{real, $\chi_\text{ee}, \chi_\text{mm}$}
%\psfrag{d}[c][c][0.8]{imaginary, $\chi_\text{ee}, \chi_\text{mm}$}}
\label{Fig:PuckFieldsBal1}}\\
\subfloat[]{
\includegraphics[width=0.9\columnwidth]{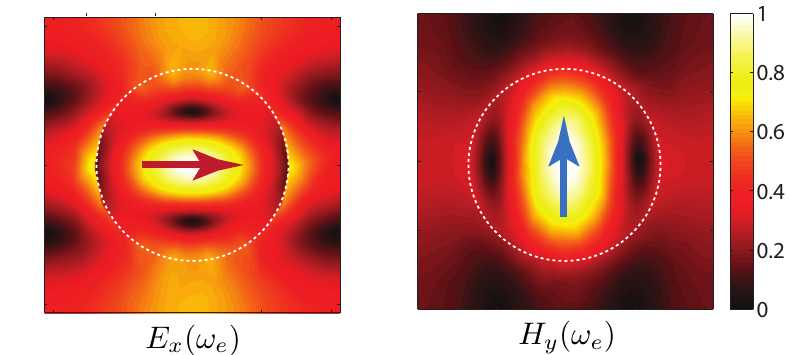}
%\psfragfig*[width=0.9\columnwidth]{BalFields}{
%\psfrag{a}[c][c][0.9]{${E_x}(\omega_{e})$}
%\psfrag{b}[c][c][0.9]{${H_y}(\omega_{e})$}}
\label{Fig:PuckFieldsBal2}}
\caption{Simulated matched metasurface. (a)~Top, transmission (solid blue line) and reflection (dashed red line) coefficients on the left and, transmission phase on the right. Bottom, real part of the retrieved electric (red line) and magnetic (blue line) susceptibilities on the left and corresponding imaginary parts on the right. (b) Electric and magnetic field distributions in the electric and magnetic resonances, respectively. All results are computed using HFSS.}
\label{Fig:PuckFieldsBal}
\end{center}
\end{figure}
The simulation results, for an unmatched metasurface,  are presented in Fig.~\ref{Fig:PuckFieldsUnBal} where the transmission and reflection coefficients and the electric and magnetic susceptibilities are plotted versus frequency on the top and bottom of Fig.~\ref{Fig:PuckFieldsUnBal1}, respectively. Note that the susceptibilities are obtained by reversing relations~\eqref{Eq:RT} and, thus, the susceptibilities are expressed in terms of the reflection and transmission coefficients~\cite{Metasurface_Synthesis_Caloz}. In Fig.~\ref{Fig:PuckFieldsUnBal2}, the electric and magnetic fields are plotted in the dielectric at resonance frequencies $\omega_e$ and $\omega_m$. As can be seen, the electric and magnetic resonances of the resonators do not overlap at the same frequency and their respective dispersions are globally different (both real and imaginary parts) thereby not satisfying the matching property. This leads to an unmatched structure (bad transmission over the considered frequency band) and unequal electric and magnetic dipolar fields. Moreover, the transmission phase exhibits two successive $\pi$ phase shift around the corresponding electric and magnetic resonances.
\begin{figure}[htbp]
\begin{center}
\subfloat[]{
\includegraphics[width=0.77\columnwidth]{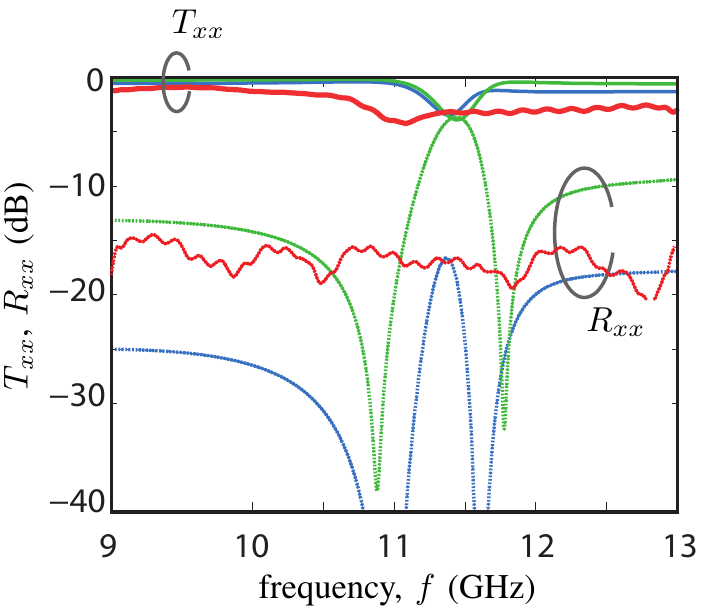}
%\psfragfig*[width=0.8\columnwidth]{Meas}{
%\psfrag{a}[c][c][1]{frequency, $f$ (GHz)}
%\psfrag{d}[c][c][1]{$T_{xx}$, $R_{xx}$ (dB)}
%\psfrag{j}[l][c][1]{$T_{xx}$}
%\psfrag{k}[l][c][1]{$R_{xx}$}}
}\\
\subfloat[]{
\includegraphics[width=0.75\columnwidth]{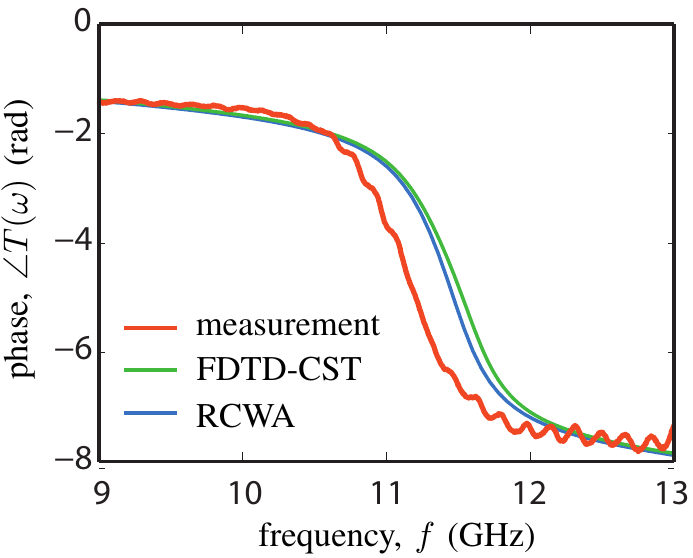}
%\psfragfig*[width=0.79\columnwidth]{Phase}{
%\psfrag{a}[c][c][1]{frequency, $f$ (GHz)}
%\psfrag{h}[c][c][1]{phase, $\angle T(\omega)$ (rad)}
%\psfrag{c}[l][c][1]{measurement}
%\psfrag{e}[l][c][1]{FDTD-CST}
%\psfrag{g}[l][c][1]{RCWA}}
}
\caption{Transmission magnitude and phase of the metasurface in Fig.~\ref{Fig:Setup}. Full-wave results are also shown for comparison, computed using CST Microwave Studio and Rigorous Coupled-Wave Analysis Technique \cite{RCWA}, where an infinite 2D array is assumed.}
\label{Fig:Measurement}
\end{center}
\end{figure}

In Fig.~\ref{Fig:PuckFieldsBal}, the structure has been tuned such that the electric and magnetic resonances almost overlap at the same frequency ($\omega_e \approx \omega_m$), the new radius of the resonators is $r = 1.4$~mm. In this latter case, the transmission is almost flat over a large bandwidth (with $|T(\omega)|\approx 100\%$). Note that the shape of the electric and magnetic dispersions are slightly different meaning that the conditions $\chi_\text{ee}(\omega) = \chi_\text{mm}(\omega)$ is not achieved here. Because of this different dispersion, the best transmission response is obtained when the two resonances are not exactly at the same frequency but they are still close enough such that the transmission phase is the sum of both electric and magnetic contributions and thus exhibits a $2\pi$ phase change.\\

\begin{figure}[htbp]
\centering
\subfloat[]{\raisebox{3mm}{
\includegraphics[width=1\linewidth]{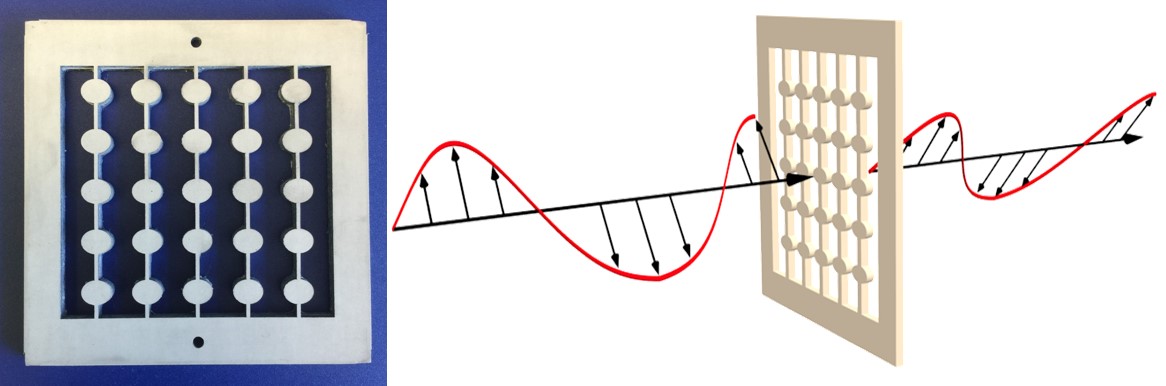}
}\label{Fig:HWP1}}\\
\subfloat[]{
\includegraphics[width=0.8\linewidth]{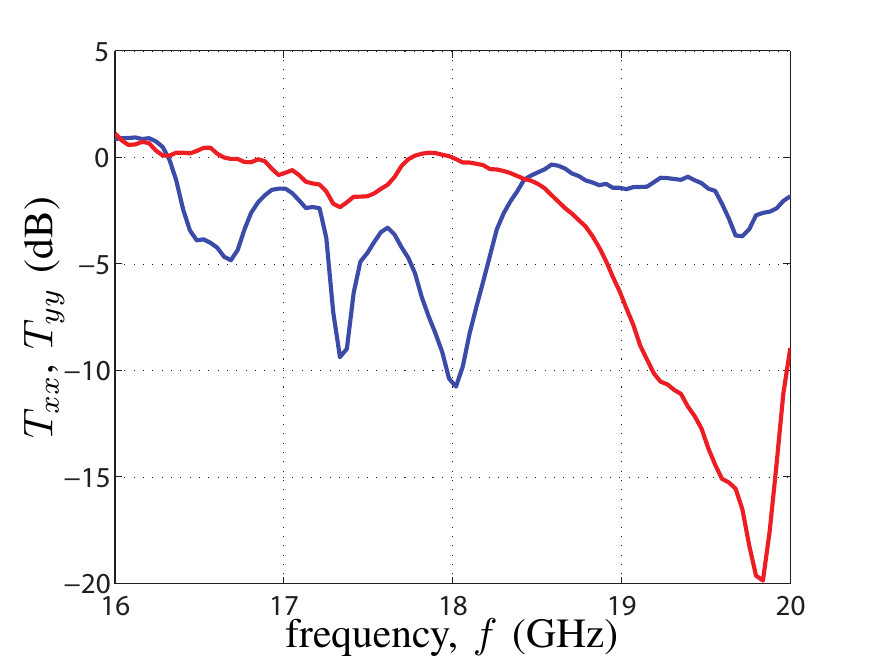}
%\psfragfig*[width=\linewidth]{HWP_T}{
%\psfrag{x}[][][1.2]{frequency, $f$ (GHz)}
%\psfrag{y}[][][1.2]{$T_{xx}$, $T_{yy}$ (dB)}}
\label{Fig:HWP2}}\\
\subfloat[]{
\includegraphics[width=0.8\linewidth]{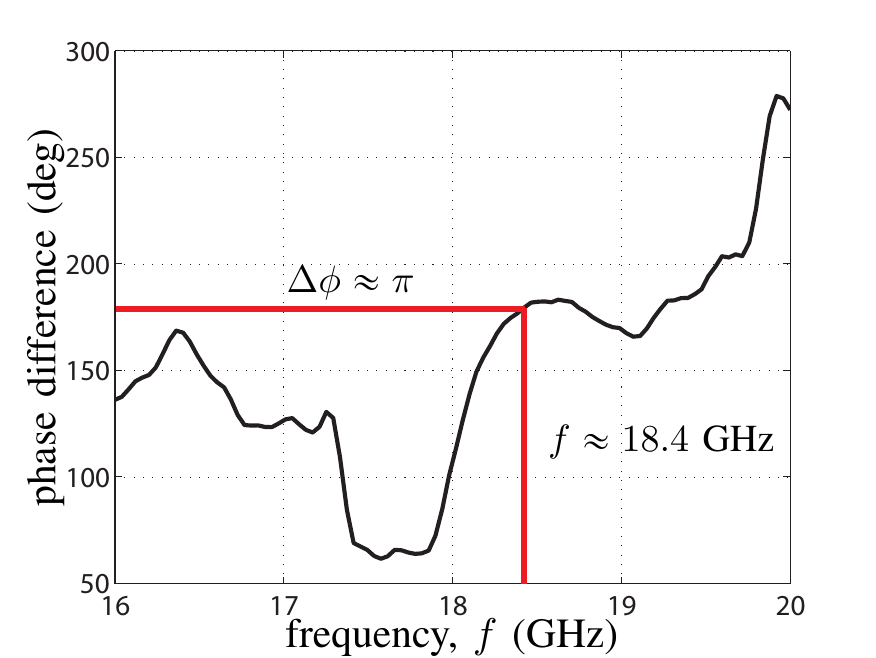}
%\psfragfig*[width=\linewidth]{HWP_phase_diff}{
%\psfrag{a}[][][1.1]{$\Delta\phi\approx\pi$}
%\psfrag{b}[l][l][1.1]{$f\approx 18.4$ GHz}
%\psfrag{x}[][][1.2]{frequency, $f$ (GHz)}
%\psfrag{y}[][][1.2]{phase difference (deg)}}
\label{Fig:HWP3}}
\caption{Half-wave plate dielectric metasurface (a) picture of the structure and representation of its rotation of polarization property. Measurements for (b) normalized transmitted power $T_{xx}$ (blue line) and $T_{yy}$ (red line), and (c) phase difference between the two polarizations.}
\label{Fig:HWP}
\end{figure}
After presenting the simulations of the metasurface without dielectric connections holding the resonators together (Figs.~\ref{Fig:PuckFieldsUnBal} and~\ref{Fig:PuckFieldsBal}), we are now interested in the more practical design presented in Fig.~\ref{Fig:FabTech}. The fabricated structure, that is shown in Fig.~\ref{Fig:Setup}, has a thickness of $t=5.08$ mm (made of 4 1.27-mm thick dielectric slabs of permittivity $\epsilon_r=10.2$ and $\tan{\delta}=0.0027$), a resonator radius of $r=4.95$ mm, a unit cell period of $P=20$ mm and a connection width of $w=2.54$ mm. The structure has been first simulated using CST Microwave studio as well as a home-made Rigorous Coupled-Wave Analysis (RCWA) code~\cite{RCWA,rumpf2012simple}, and it has also been experimentally measured using the setup in Fig.~\ref{Fig:Setup}. The simulated and measured reflection and transmission coefficients as well as the transmission phase are plotted in Fig.~\ref{Fig:Measurement} for $x$ polarization (orthogonal to the resonator connections). As can be seen, the metasurface is well matched and presents a relatively good and flat transmission over the measured bandwidth. A -0.6~dB dip in the transmission occurs at the resonance frequency which, as explained above, is mostly due to dielectric loss and, to a lesser extent, the unequal dispersion of the electric and magnetic resonances. There is small down-frequency shift ($\approx 3\%$) between the measured and the simulated results that may be explained by slight dimension variations due to the fabrication process. The transmission phase variation, which covers a $2\pi$ range, is an indication that the electric and magnetic resonances overlap at the same frequency ($\omega_e\approx\omega_m$).

\section{Monochromatic Waveplates Applications}

In the previous section, we have demonstrated the capability of the proposed structure to control the phase while maintaining a good transmission efficiency over a large bandwidth. Now, this concept is used to implement two monochromatic waveplates, namely a half-wave plate and a quarter-wave plate. Wave-plates are birefringent structures presenting different phase shifts to $x$ and $y$ polarizations. A half-wave plate introduces a phase shift difference of $\pi$ and, thus, rotates a linear polarization by $90^\circ$ or changes the handedness of circularly polarized waves. A quarter-wave plate introduces a phase shift difference of $\pi/2$ and is consequently a linear-to-circular polarization converter~\cite{Saleh_Teich_FP}.

\begin{figure}[htbp]
\centering
\subfloat[]{\raisebox{3mm}{
\includegraphics[width=1\linewidth]{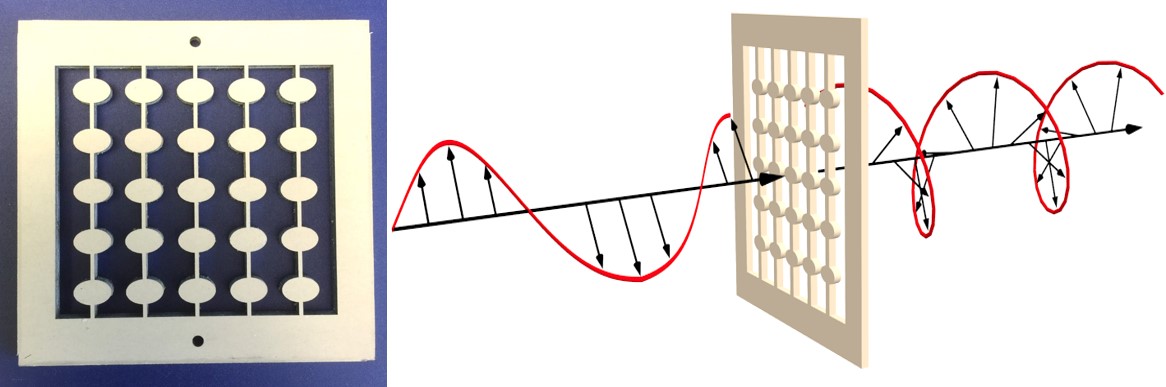}
\label{Fig:QWP1}}}\\
\subfloat[]{
\includegraphics[width=0.8\linewidth]{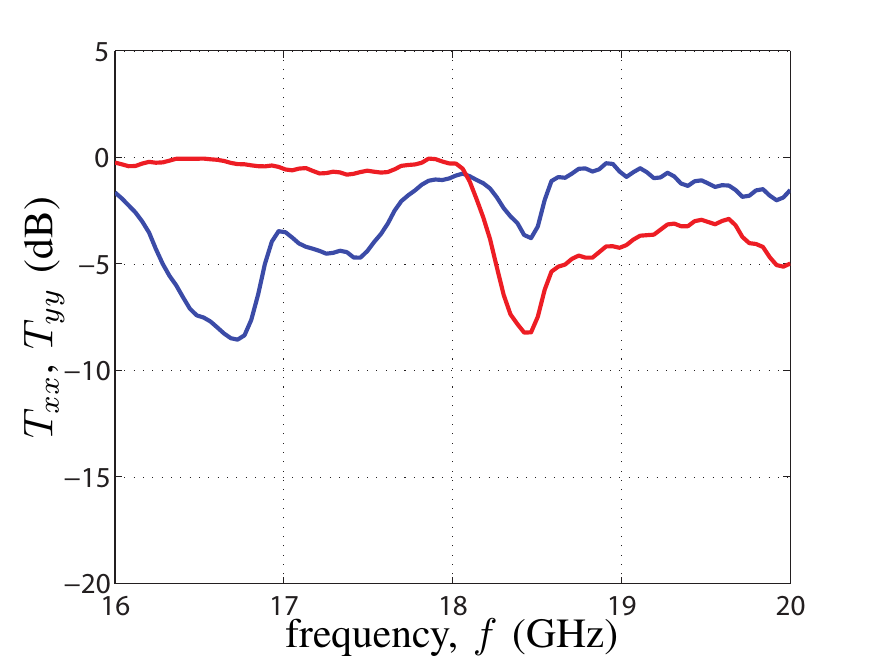}
%\psfragfig*[width=\linewidth]{QWP_T}{
%\psfrag{x}[][][1.2]{frequency, $f$ (GHz)}
%\psfrag{y}[][][1.2]{$T_{xx}$, $T_{yy}$ (dB)}}
\label{Fig:QWP2}}\\
\subfloat[]{
\includegraphics[width=0.8\linewidth]{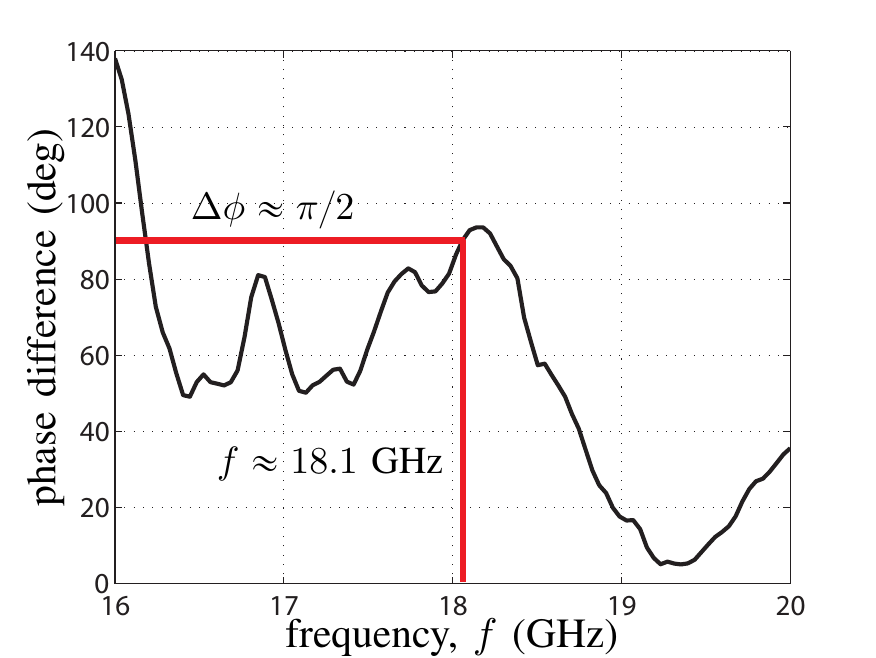}
%\psfragfig*[width=\linewidth]{QWP_phase_diff}{
%\psfrag{a}[][][1.1]{$\Delta\phi\approx\pi/2$}
%\psfrag{b}[][][1.1]{$f\approx 18.1$ GHz}
%\psfrag{x}[][][1.2]{frequency, $f$ (GHz)}
%\psfrag{y}[][][1.2]{phase difference (deg)}}
\label{Fig:QWP3}}
\caption{Quarter-wave plate dielectric metasurface (a) picture of the structure and representation of its linear-to-circular polarization conversion property. Measurements for (b) normalized transmitted power $T_{xx}$ (blue line) and $T_{yy}$ (red line), and (c) phase difference between the two polarizations.}
\label{Fig:QWP}
\end{figure}

In order to control the $x$ and $y$ polarizations differently, we have used dielectric resonators with elliptical shapes. Elliptical resonators provide extra degrees of freedom to tune the phase shifts of both polarizations. In the two structures discussed thereafter, the frequency of operation is 18~GHz, the unit cell size is $14.9$~mm, the thickness is $5.08$~mm ($\approx \lambda_0/3.3$), the width of the connections is $1.25$~mm and the dielectric substrates are Rogers 3210 with $\epsilon_r = 10.2$ and $\tan{\delta}=0.0027$. The ellipses are defined with respect to their major axis ($a$) and minor axis ($b$) that are in the $x$ and $y$ direction, respectively. Note that the dielectric connections holding the resonators are in the $y$ direction, as illustrated in Fig.~\ref{Fig:Setup}.

The first waveplate that was realized, is the half-wave plate that is shown in Fig.~\ref{Fig:HWP1}. It was tuned such that $x$ and $y$ polarizations are transmitted with equal amplitudes (and ideally full transmission) but with a phase shift difference, $\Delta \phi=|\phi_x-\phi_y|$, of $\Delta \phi=\pi$. The resonators axes are $a = 4.5$ mm and $b=3.7$ mm. The measured transmitted power is plotted in Fig.~\ref{Fig:HWP2} for the $x$ and $y$ polarizations. It can be seen that the two transmission curves overlap at about 18.4~GHz which corresponds to a transmission of -1 dB. The phase difference, $\Delta\phi$, is plotted in Fig.~\ref{Fig:HWP3}. At 18.4~GHz, the phase difference is, as required, $\Delta\phi\approx \pi$. Therefore, the metasurface behaves as a half-wave plate at this specific frequency. Note that the transmission coefficients in Fig.~\ref{Fig:HWP2} have been normalized with respect to the exciting horn antenna reference transmission coefficients. The reason why $T_{xx}$ and $T_{yy}$ are above 0~dB on the left-hand side of Fig.~\ref{Fig:HWP2} is that the metasurface scatters more into the measuring area than the reference antenna does.

The second waveplate is the quarter-wave plate shown in Fig.~\ref{Fig:QWP1}. As for the half-wave plate, the metasurface must transmit the $x$ and $y$ polarizations with equal amplitudes but with a phase shift difference of $\Delta \phi=\pi/2$. The results are shown in Figs.~\ref{Fig:QWP2} and~\ref{Fig:QWP3} for the transmitted power of $x$ and $y$ polarizations and the phase difference, respectively. For this structure, the two transmission curves overlap at 18.1~GHz for a transmission of about -1 dB and a $\Delta\phi\approx\pi/2$. At this frequency, the metasurface effectively behaves as a quarter-wave plate.

The fact of using elliptical resonators deteriorates the frequency dependent transmission response of the metasurface (Figs.~\ref{Fig:HWP} and~\ref{Fig:QWP}) when compared to response of the cylindrical resonator metasurface (Fig.~\ref{Fig:Measurement}) which has a very flat transmission profile over the bandwidth of interest. Consequently, elliptical resonators offer better control of $x$ and $y$ polarizations at the cost of a smaller transmission bandwidth. It is also apparent that the transmission profiles of $T_{xx}$ and $T_{yy}$ are very different, $T_{yy}$ being much more affected by the presence of the connections than $T_{xx}$.

\section{Conclusion}

We have proposed a very simple dielectric metasurface design working at microwave frequencies. The structure is made of dielectric resonators held together by dielectric connections for mechanical support. The resonators, which in the simplest cases have a cylindrical shape, present electric and magnetic resonances which can be tuned to control the transmission response of the metasurface. Broadband transmission can consequently be achieved as well as monochromatic wave transformation. A more sophisticated control of the transmission can be obtained by using more complex shapes for the resonators such as, for instance, elliptical resonators.

In order to demonstrate the capabilities of the proposed design, three metasurfaces were realized. The first metasurface, made of cylindrical resonators, has a broadband response with a $2\pi$ phase shift around the resonance frequency. The two other metasurfaces, made of elliptical resonators, have a monochromatic response and control transmission for $x$ and $y$ polarizations. They perform the operation of a half-wave plate and a quarter-wave plate, respectively. In all cases, the transmission efficiency remained almost always around $90\%$, which makes this metasurface design an alternative approach to the realization of microwave metasurfaces usually made with metallic scattering particles.

\section*{Acknowledgment}
This work was accomplished in the framework of the Collaborative Research and Development Project CRDPJ 478303-14 of the Natural Sciences and Engineering Research Council of Canada (NSERC) in partnership with the company Metamaterial Technology Inc. It was also partly funded by the Deanship of Scientific Research (DSR), King Abdulaziz University, under grant No. 25-135-35-HiCi.

\bibliographystyle{IEEETran}
\bibliography{Achouri_Dielectric_Metasurface_2016}

\end{document}